\documentclass{iaus}
\usepackage{graphicx}

\title[Search for cool giant exoplanets combining coronagraphy and differential imaging] 
{Search for cool extrasolar giant planets combining coronagraphy, spectral and angular differential imaging}

\author[A.-L. Maire et al.]   
{Anne-Lise Maire$^1$, Anthony Boccaletti$^2$, Julien Rameau$^3$, Ga\"el Chauvin$^3$, Anne-Marie Lagrange$^3$, Micka\"el Bonnefoy$^4$, Silvano Desidera$^5$, M\' elody Sylvestre$^2$, Pierre Baudoz$^2$, Rapha\"el Galicher$^2$, \and David Mouillet$^3$}

\affiliation{$^1$LUTH, Obs. Paris/CNRS/Univ. Paris 7, 92195 Meudon, France \\ email: {\tt anne-lise.maire@obspm.fr};
$^2$LESIA, Obs. Paris/CNRS/Univ. Paris 6/Univ. Paris 7, 92195 Meudon, France;
$^3$Institut de Plan\'etologie et d'Astrophysique de Grenoble, University Joseph Fourier/CNRS, 38041 Grenoble, France;
$^4$Max Planck Institute for Astronomy, D-69117 Heidelberg, Germany;
$^5$INAF - Osservatorio Astronomico di Padova, I-35122 Padova, Italy}

\pubyear{2013}
\volume{299}  
\pagerange{xxx}
\jname{Exploring the formation and evolution of planetary systems}
\editors{M. Booth, B.~C. Matthews, \& J.~R. Graham, eds.}

\newcommand\fref{$F_{\rm{r}}$ } 
\newcommand\fsubtrac{$F_{\rm{s}}$ } 

\begin{document}

\maketitle

\begin{abstract}
Spectral differential imaging (SDI) is part of the observing strategy of current and on-going high-contrast imaging instruments on ground-based telescopes. Although it improves the star light rejection, SDI attenuates the signature of off-axis companions to the star, just like angular differential imaging (ADI). However, the attenuation due to SDI has the peculiarity of being dependent on the spectral properties of the companions. To date, no study has investigated these effects. Our team is addressing this problem based on data from a direct imaging survey of 16 stars combining the phase-mask coronagraph, the SDI and the ADI modes of VLT/NaCo. The objective of the survey is to search for cool (Teff$<$1000–-1300~K) giant planets at separations of 5–-10~AU orbiting young, nearby stars ($<$200~Myr, $<$25~pc). The data analysis did not yield any detections. As for the estimation of the sensitivity limits of SDI-processed images, we show that it requires a different analysis than that used in ADI-based surveys. Based on a method using the flux predictions of evolutionary models and avoiding the estimation of contrast, we determine directly the mass sensitivity limits of the survey for the ADI processing alone and with the combination of SDI and ADI. We show that SDI does not systematically improve the sensitivity due to the spectral properties and self-subtraction of point sources. 
\end{abstract}

\firstsection 
\section{Introduction}

Most of the direct imaging searches for young giant exoplanets conducted so far use adaptive optics (AO) combined with broad-band imaging in the near-infrared (for instance, \cite{Lafreniere2007, Chauvin2010, Vigan2012}). \cite{Biller2007} instead performed a survey using narrow-band adaptive optics imaging and spectral differential imaging (SDI, \cite{Racine1999}). More recently, the NICI team on Gemini (\cite{Liu2010}) and our team on VLT/NaCo (Maire et al., in prep.) completed surveys to search for cool ($<$1\,300~K) and close-in ($>$5--10~AU) young giant exoplanets combining several state-of-the-art high-contrat imaging techniques: coronagraphy, SDI, and angular differential imaging (ADI, \cite{Marois2006}). This observing strategy is similar to the strategy which will be employed for SPHERE (\cite{Beuzit2008}) and GPI (\cite{Macintosh2008}). \cite{Biller2007} and \cite{Nielsen2008} determined SDI sensitivity limits by converting the contrast levels measured in the reduced images into masses using evolutionary models. In this proceeding, we show that this method is optimistic for the analysis of SDI-processed images and instead we propose a straightforward method based on the injection of synthetic planets in the raw data, combined with the use of evolutionary models. This framework should serve as a basis for the data reduction and analysis of SPHERE and GPI. We present the observing strategy, the procedure used to analyze and interprete the data, and the detection limits of the survey.

\section{Observing strategy}

\begin{figure}
\centering
\includegraphics[trim = 22mm 5mm 4.5mm 11mm, clip,width=.45\textwidth]{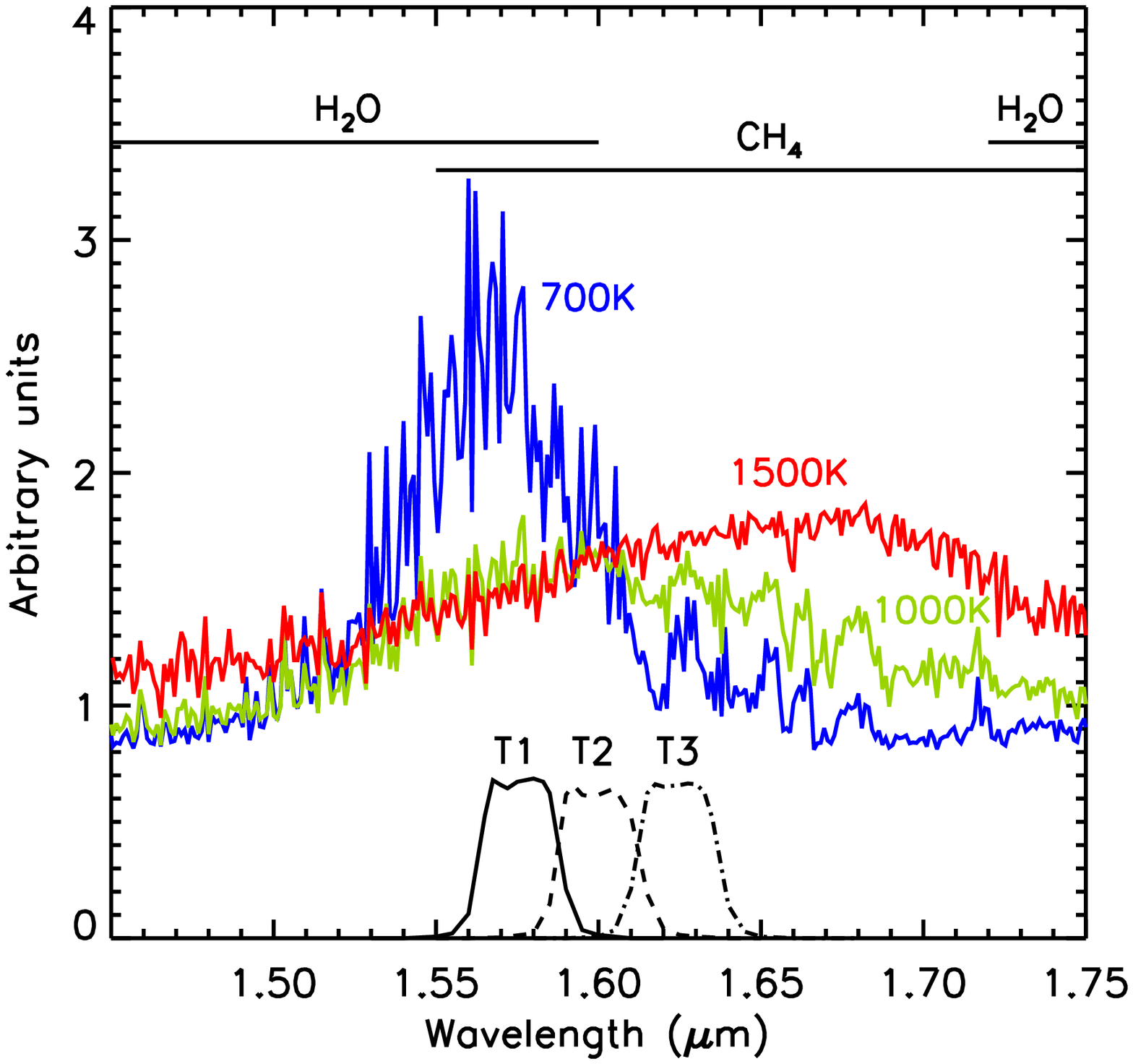}
\includegraphics[trim = 20mm 5mm 7.5mm 11mm, clip, width=.45\textwidth]{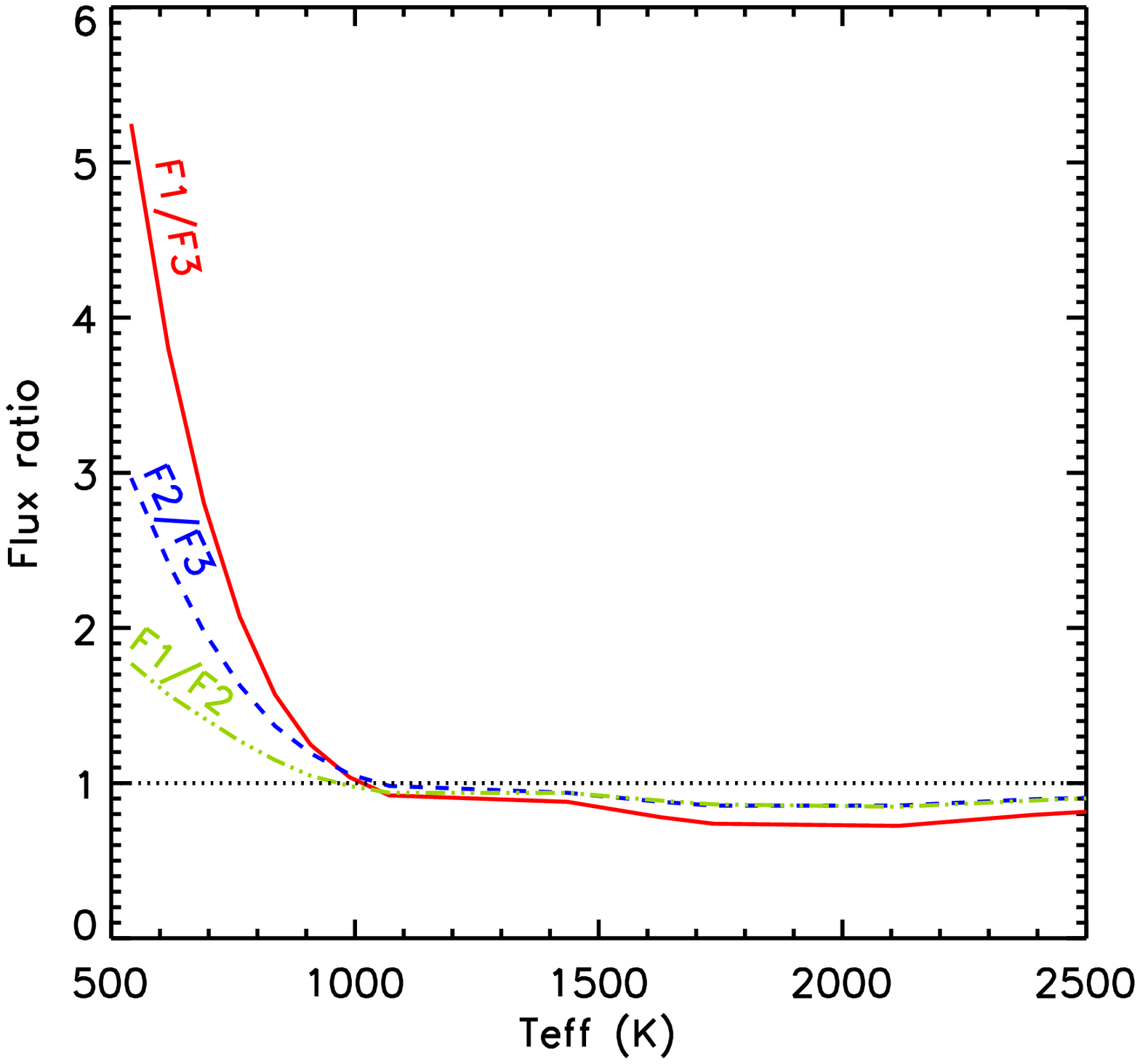}
\caption{Left: Spectra of model atmospheres of giant planets for different effective temperatures (colored solid lines, BT-Settl models from \cite{Allard2011}). Each spectrum is normalized to its value at 1.6~$\mu$m and is vertically shifted by a constant. The transmission of the three SDI filters of NaCo are shown (black curves). Theoretical absorption bands of water and methane are also indicated. Right: Flux ratio as a function of the effective temperature derived for the NaCo SDI filters. $F_1$, $F_2$, and $F_3$ refer to the fluxes in the filters at 1.575, 1.6, and 1.625~$\mu$m.}
\label{fig:spectracoldegp}
\end{figure}

We use an observing strategy combining the four-quadrant phase mask, the ADI and SDI modes of VLT/NaCo. SDI exploits the natural wavelength dependence of a star image (\cite{Racine1999}). Two simultaneous images taken at different wavelengths are rescaled spatially to correct for the PSF chromaticity and rescaled in intensity to correct for the filter transmission before being subtracted. SDI is intended to take advantage of the presence of a methane absorption band at $\sim$1.6~$\mu$m in the spectra of cool ($<$1\,300~K) giant planets (Fig.~\ref{fig:spectracoldegp}, left), that is not present in the star's spectrum. Based on a complete compilation of young and nearby stars recently identified in young co-moving groups and from systematic spectroscopic surveys, we selected a sub-sample of 16 stars, mostly AFGK spectral types, according to their age ($\lesssim$200~Myr), distance ($\lesssim$25~pc) and R-band brightness ($\lesssim 9.5$). The age cut-off ensures that the cool companions detected will have planetary masses. The distance cut-off guarantees that these targets are the most favorable for the detection of cool companions. The targets are brighter than $R=9.5$ in the visible to allow good AO efficiency.

\section{Analysis of spectral differential imaging data}
In practice, SDI produces a significant attenuation of the planet itself that has to be quantified to derive accurate photometry. First, this attenuation depends on the planet's spectral properties, which are determined in first approximation by the effective temperature, as illustrated in the right panel of Fig.~\ref{fig:spectracoldegp}. We represent the flux ratios between the NaCo SDI filters as a function of their effective temperature. For temperatures higher than the methane condensation temperature (1\,300~K), the flux ratios are close to 1, so the flux differences are minimum and the self-subtraction of point sources is high. Interestingly, the flux ratios remain close to 1 below 1\,300~K down to 1\,000~K. Thus, the SDI domain is constrained to temperatures lower than those what we expect from the methane condensation temperature. Below 1\,000~K, the flux ratios increase and reach the highest values for the $F_1/F_3$ ratio. Consequently, we consider for the data analysis only the image subtraction $I_1 - I_3$, with $I_1$ ($I_3$) is the image at 1.575 (1.625)~$\mu$m. The second factor to account for when deriving the SDI attenuation is the geometrical overlap between the two planet images in $I_1$ and $I_3$ due to the spatial rescaling. This attenuation is maximum near the image center and decreases with angular separation.

In the general context of high-contrast imaging with a single spectral filter, the detection limits are measured on contrast maps. In most cases, we use the azimuthal standard deviation to derive a 1-dimensional contrast plot, which after correction from various attenuations (ADI and/or coronagraph) is then converted into mass limits at 5~$\sigma$ according to a given evolutionary model. The problem with SDI is different as we measure a differential intensity. It can be expressed as follows (no coronagraph and no ADI): 

\begin{equation}
F_{\rm{rs}}=F_{\mathrm{r}}-F_{\mathrm{s}} \times \alpha \times \phi(\vec{r})
\label{eq:fsdi}
\end{equation}
 
where \fref and \fsubtrac are the object intensities at respectively $\lambda_{\rm{r}}$ and $\lambda_{\rm{s}}$, $\alpha$ the intensity rescaling factor, $\phi(\vec{r})$ the attenuation due to the spatial rescaling ($\phi \sim 1$ close to the image center and $\phi \sim 0$ at large separations) and $\vec{r}$\,=\,$(r,\theta)$. Since the residual flux of a point source can exhibit positive or negative values, like the residual  noise measured in the processed image, it becomes difficult to disentangle the noise from a planet signal. The method used for single-band surveys (ADI, coronagraphy) is no longer valid. From Eq.~(\ref{eq:fsdi}), we see that, providing $\phi(\vec{r})$ is calibrated (by using the point-spread functions measured in the different filters), we have to test the individual intensities \fref and \fsubtrac for all planet masses (taken from an evolutionary model) which reproduce the measured $F_{\rm{rs}}$. We expect that several values of \fref and \fsubtrac can match the observations, resulting in degeneracies in planet mass. The number of degeneracies and the values of the mass solutions depend on $\phi(\vec{r})$, i.e. the position in the image field. To break these degeneracies, we need to compare the SDI sensitivity limits to the limits derived in broad-band imaging. For the analysis of our survey, we consider and compare two processings: classical ADI (noted ADI, \cite{Marois2006}) and an algorithm combining SDI as a first step and classical ADI as a second step (refered as ASDI). 

\section{Sensitivity limits of the survey}

The data analysis did not yield any detections. We focus instead on the interpretation of the detection limits. For this, we applied a straightforward and robust method based on the injection of fake planets in the data set at the cost of a longer computing time and sparsity in the detection map. The use of fake planets assumes that the PSF is the same in the coronagraphic (or saturated) and the out-of mask (or unsaturated) images. This hypothesis strongly depends on the AO-loop and photometric stabilities, and we should expect variations from one data set to another. The fake planets are injected simultaneously at several separations and position angles in the raw datacubes with the flux predicted by the BT-Settl models (\cite{Allard2011}) and processed with ADI and ASDI. For each separation, we measured the planet fluxes using aperture photometry and averaged them. The noise was measured on the data processed without the fake planets in rings of 1~FWHM and scaled to the same aperture size. The process was repeated for each model mass. Finally, we interpolate for each separation the signal-to-noise ratio vs. mass relation to derive the corresponding mass achieved at 5~$\sigma$.

\begin{figure}
    \centering
	\includegraphics[trim = 15mm 5.5mm 9mm 11mm, clip, width=.45\textwidth]{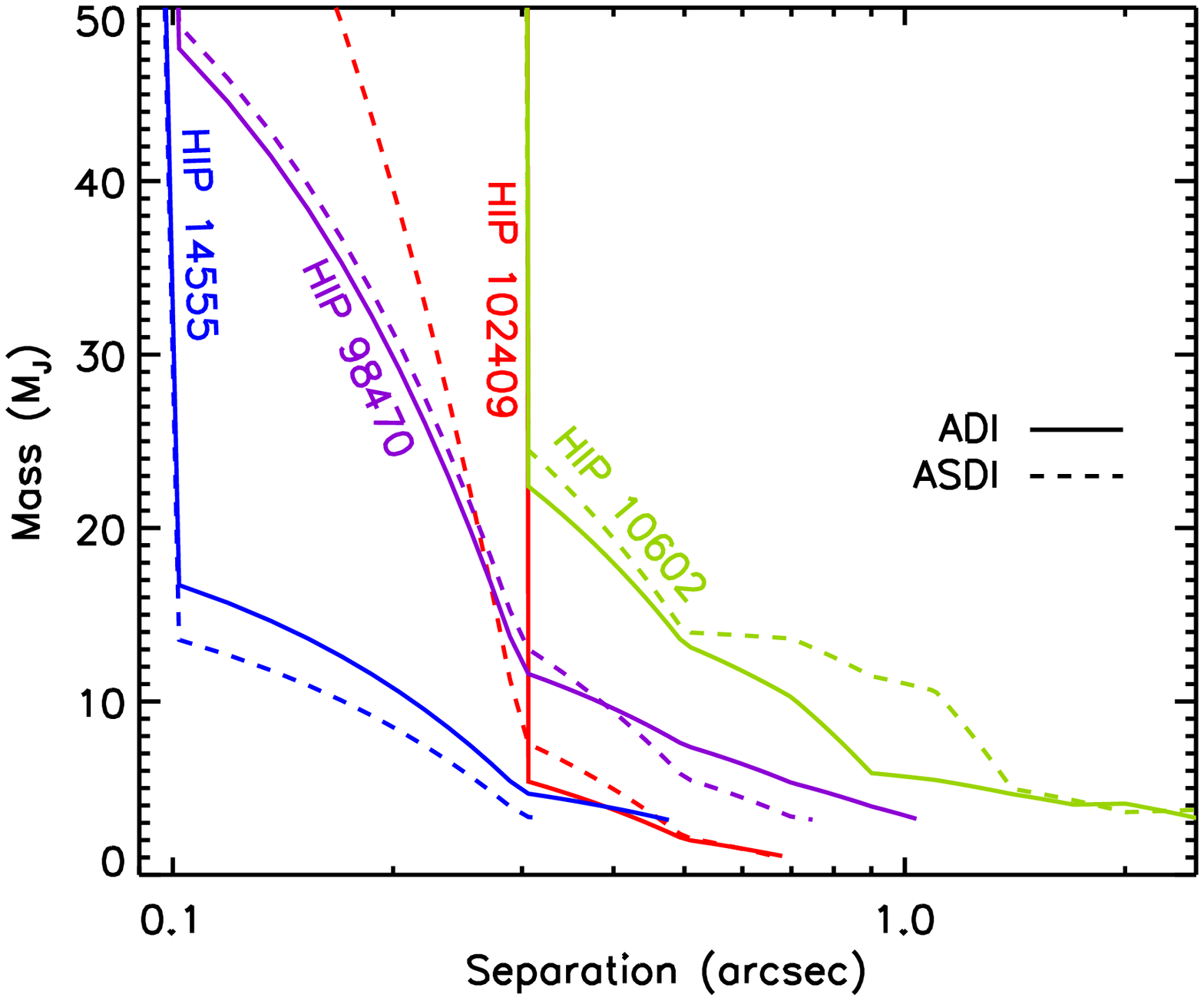}
	\includegraphics[trim = 15mm 5.5mm 9mm 11mm, clip, width=.45\textwidth]{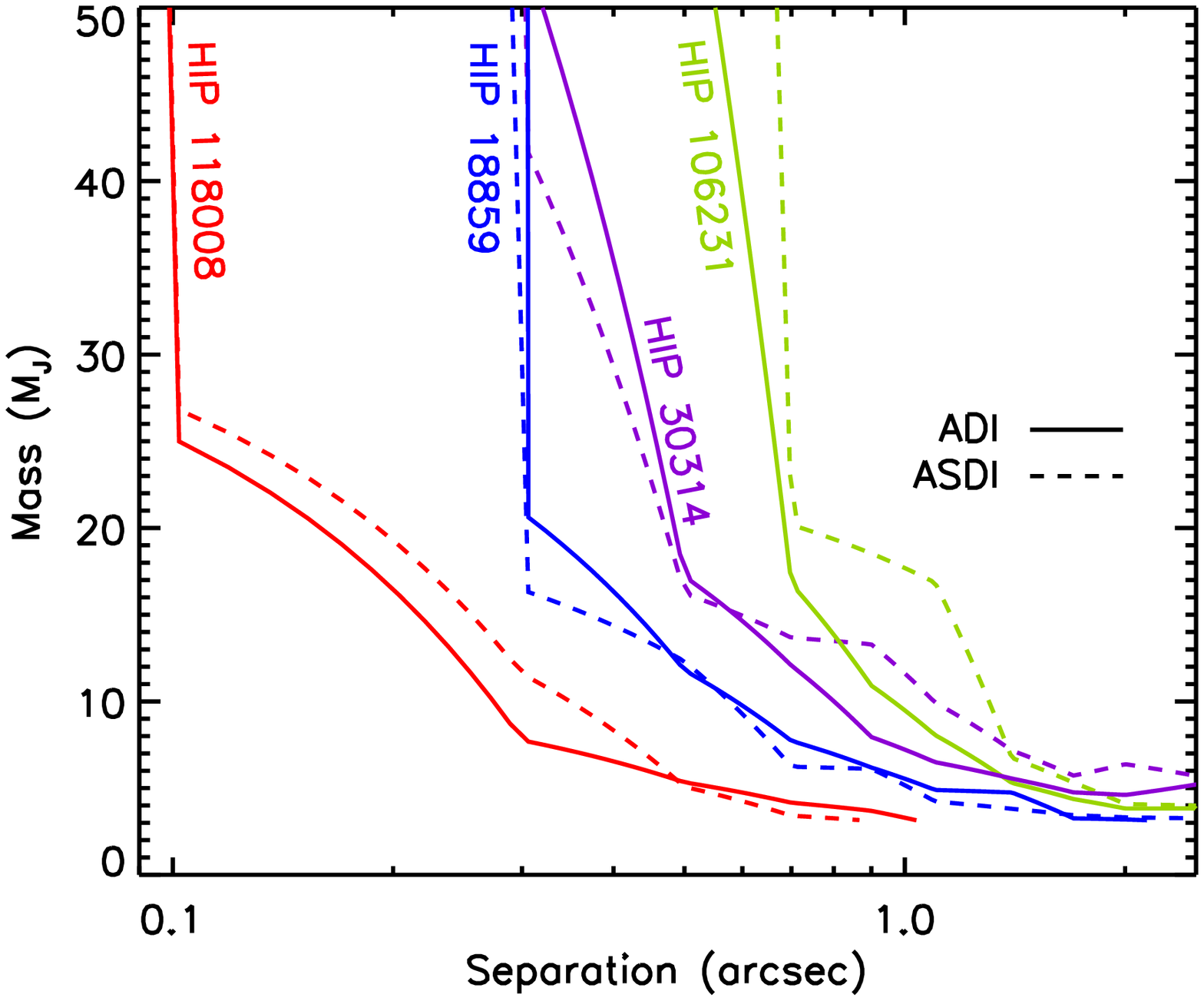}
\caption{Sensitivity limits at 5 $\sigma$ of the survey in Jupiter masses of detectable companions (BT-Settl models, \cite{Allard2011}) for the youngest stars of the survey ($\leq$70~Myr). The curves are derived for two pipelines: classical ADI (solid curves) and ASDI (dashed curves).}
\label{fig:senslim}
\end{figure}

Figure~\ref{fig:senslim} presents the mass sensitivity limits at 5~$\sigma$ of the survey as a function of the angular separation for the youngest stars of the survey ($\leq$70~Myr). The detection limits are cut when the minimum effective temperature available in the BT-Settl grids ($\sim$500~K) is reached. The main result is that SDI can either improve or degrade the sensitivity (ASDI curve reaching either lower or larger masses than the corresponding ADI curve). There are no particular trends with the star age and the angular separation. We are analyzing possible trends with parameters representative of the quality of the observations.

\section{Conclusion}
We show that the interpretation of SDI-processed images requires a different analysis than that used for broad-band direct imaging surveys. In particular, we need to compare the sensitivity limits to those obtained in broad-band imaging in order to break the degeneracies in planet mass. This analysis demonstrates that SDI does not improve systematically the sensitivity because of the spectral properties and self-subtraction of point sources. SDI on NaCo gives the best performance for separations of 0.5--1$''$. We expect improved performances with SPHERE and GPI thanks to their extreme AO systems.

\begin{discussion}

\discuss{Graham}{How does SDI improve with increased spectral resolution?}

\discuss{Maire}{When you increase the spectral resolution, the planet signals in the spatially rescaled images will be less separated (the spatial rescaling factor is proportional to the wavelength ratio). As a result, the self-subtraction of the planet will be higher and the SDI performance will be worse.}
\end{discussion}

\end{document}